\pdfoutput = 1
\documentclass[conference]{IEEEtran}
\IEEEoverridecommandlockouts
\renewcommand{\baselinestretch}{0.87}

\usepackage{cite}
\usepackage{float}
\usepackage{url}
\usepackage{amsmath,amssymb,amsfonts}
\usepackage{algorithmic}
\usepackage{graphicx}
\usepackage{textcomp}
\usepackage{xcolor}

\def\BibTeX{{\rm B\kern-.05em{\sc i\kern-.025em b}\kern-.08em
    T\kern-.1667em\lower.7ex\hbox{E}\kern-.125emX}}
\begin{document}

\title{End-to-end 100-TOPS/W Inference With Analog In-Memory Computing: Are We There Yet?\\
\thanks{We thank Abu Sebastian for fruitful discussions. This work was supported in part by the Italian Ministry for Education, University and Research (MIUR) under the program “Dipartimenti di Eccellenza (2018-2022)" and WiPLASH (Wireless Plasticity for Heterogeneous Massive Computer Architectures) Grant Agreement No. 863337 funded from the European Union’s Horizon 2020 research and innovation program.}}
\author{\IEEEauthorblockN{Gianmarco Ottavi\IEEEauthorrefmark{1}, Geethan Karunaratne\IEEEauthorrefmark{2}, Francesco Conti\IEEEauthorrefmark{1}, Irem Boybat\IEEEauthorrefmark{2}, Luca Benini\IEEEauthorrefmark{3}\IEEEauthorrefmark{1} and Davide Rossi\IEEEauthorrefmark{1}} 
\IEEEauthorblockA{\textit{DEI, University of Bologna, Italy}\IEEEauthorrefmark{1} \quad \textit{IBM Research Europe}\IEEEauthorrefmark{2} \quad \textit{IIS lab, ETH Zurich, Switzerland}\IEEEauthorrefmark{3} \\
$\{$gianmarco.ottavi2, davide.rossi, f.conti$\}$@unibo.it
$\{$lbenini$\}$@iis.ee.ethz.ch
$\{$kar, ibo$\}$@zurich.ibm.com}}

%\author{Omitted for Blind Review}

%\author{\IEEEauthorblockN{1\textsuperscript{st} Gianmarco Ottavi}
%Bologna, Italy \\
%\and
%\and
%\and
%\and
%\and
%%City, Country \\

\maketitle

\begin{abstract}
%Addressing the fast-growing computational requirements of Deep Neural Networks (DNN) inference within tight power/area constraints is one of the main challenges for embedded Artificial Intelligence (AI).
%Analog In-Memory Computing (AIMC) is a promising technology to accelerate DNN workloads dominated by multiply-accumulate operations, thanks to a high degree of parallelism and energy-efficient analog operation (up to hundreds of Top/s/W at peak).
In-Memory Acceleration (IMA) promises major efficiency improvements in deep neural network (DNN) inference, but challenges remain in the integration of IMA within a digital system. We propose a heterogeneous architecture coupling 8 RISC-V cores with an IMA in a shared-memory cluster, analyzing the benefits and trade-offs of in-memory computing on the realistic use case of a MobileNetV2 bottleneck layer. We explore several IMA integration strategies, analyzing performance, area, and energy efficiency. We show that while pointwise layers achieve significant speed-ups over software implementation, on depthwise layer the inability to efficiently map parameters on the accelerator leads to a significant trade-off between throughput and area. We propose a hybrid solution where pointwise convolutions are executed on IMA while depthwise on the cluster cores, achieving a speed-up of 3x over SW execution while saving 50\% of area when compared to an all-in IMA solution with similar performance.
% 
% While, accelerating standard Convolutional Neural Network (CNN) layers IMA provides up to 36x better performance when compared to the 8-cores software solution, when employed on bottlenecks the total speed-up drops to 3x when maximising for an area efficient solution.
\end{abstract}


\section{Introduction}

%\subsection{In-memory computing}
Analog In-Memory Computing (AIMC) is an emerging paradigm holding promise to overcome the well-known von Neumann bottleneck by executing operations such as matrix-vector products in the analog domain within a crossbar arrangement, with millions of operations executed simultaneously. Both charge-based memory technologies (e.g. SRAM, DRAM, and flash), and resistance-based memory technologies (e.g. RRAM, PCM, and STT-MRAM) can serve as elements for such computational units \cite{sebastian2020memory}.

Among several application domains, demonstrations of AIMC-based architectures have appeared in the field of Deep Neural Network (DNN) inference acceleration, showing outstanding peak energy efficiency in the order of hundreds of TOPS/W \cite{InMComp, sebastian2020memory}. An early market industrial example is represented by Mythic \cite{mythicHC2018}, claiming efficiency of 4 TOPS/W exploiting 8-bit flash-based Mythic Analog Matrix Processors (MAMP) arranged as a systolic array, all connected through a mesh topology network on chip.
From a research perspective, several approaches claimed tens to hundreds of TOPS/W by exploiting several different approaches, with a quite diverse set of choices in levels of numerical precision and memory technologies~\cite{InMComp, sebastian2020memory}.

However, several fundamental challenges are still open to achieve the claimed levels: the intrinsic variability of analog computing both in the charge based and resistive domain \cite{sebastian2020memory}; difficulties in dealing with low-precision computations that are often the only ones supported by AIMC-based architectures \cite{sebastian2020memory}; the necessity of specialized training \cite{seb2020mixed};  the poor flexibility of IMC, that is well matched only for a limited set of primitives such as matrix-vector multiplications \cite{InMComp}.
As a result, most AIMC-based architectures fabricated so far have been demonstrated on trivial neural networks (up to ten layers) trained on single layers or simple data sets such as CIFAR-10 or MNIST \cite{sebastian2020memory}, which are not representative of real-life, DNN-based applications.

In this work, we focus on the architectural challenges described above. To tackle the limited flexibility of AIMC-based computing, some architectures couple general-purpose processors to analog in-memory computing cores. This allows extending the functionality of In-Memory Accelerators (IMA) creating heterogeneous analog/digital computing tiles, connected to the system bus \cite{mythicHC2018,jia2020programmable}. However, performing linear operators with accelerators such as IMA moves the bottleneck of the computation to the digital part.
For this reason, augmenting the heterogeneous cluster with a single core might not be sufficient to sustain the computing requirements of IMAs; moreover, low bandwidth and high communication latency between the processor and the IMA might form a remarkable bottleneck for heterogeneous computing.

We propose a new paradigm for AIMC-based heterogeneous computing, where an IMA is integrated within a parallel tightly-coupled cluster of RISC-V processors. We present a design space exploration based on a key building block of the MobileNetV2 CNN, representative for a wide range of modern DNNs leveraging depthwise convolutions to reduce the size of the model by up to one order of magnitude with respect to first-generation models. We demonstrate that, for this use-case, the proposed approach improves performance by more than one order of magnitude with respect to traditional approaches where the IMA is connected through a low-bandwidth, high-latency system bus \cite{jia2020programmable}. Finally, we analyze the remaining architectural bottlenecks for MobilenetV2 execution on such heterogeneous system. The IMA on itself can reach outstanding performance and efficiency peaks that are dictated by the size of the activated crossbar given the constant time to output for analog computations. But, the cost for auxiliary computation, data marshalling and inefficiency of depthwise layers from a significant constriction for efficiency (80\%) suggesting to further extend these clusters with specialized digital accelerators better tuned for these functions.

\section{Background}
\subsection{PCM-based In-Memory Accelerator}
\label{sec:in-mem_comp}
% In this section, we describe the basic operations of the PCM in-memory crossbar array used in this work.
The In-Memory Accelerator (IMA) used in this paper is based on a Phase-Change Memory (PCM) crossbar array \cite{hermesCor}. In this architecture, the memory devices are resistors with programmable conductance placed at the crosspoints of a 2D array with one terminal connected to horizontal wires called \textit{bitlines} and the other terminal connected to vertical wires called \textit{wordlines}, enabling execution of several computational primitives concurrently.

To perform the product of a matrix $\mathbf{A}$ by a vector $\mathbf{x}$, the PCM devices are programmed with conductance values proportional to the values $\mathbf{A}_{ij}$ of $\mathbf{A}$, with a precision of 4 bits (signed). Then the wordlines are driven with voltage pulses, whose duration are proportional to $\mathbf{x}_j$, using a set of digital-to-analog converters (DACs) with 8 bits of precision. By Ohm's law, each PCM device contributes a current proportional to $\mathbf{A}_{ij}\cdot\mathbf{x}_j$ on the $i$-th bitline, resulting in a total integrated current proportional to the dot product $\mathbf{y}_i = \sum_j \mathbf{A}_{ij}\cdot\mathbf{x}_j$. At the end of each bitline, there is an analog-to-digital converter (ADC) used to sample the bitline current and convert it into an 8-bit digital value.

For DNN inference, the $\mathbf{A}$ matrix can be used to store the weights of the linear part of a Fully Connected, Convolutional, or Depthwise Convolutional layer. Note that typically 2 PCM devices are used to denote a signed weight~\cite{Y2020joshiNatComm}. In conventional digital architectures, the dot product of 4-bit weights and 8-bit input activations require a high-precision intermediate representation (often, 32 bits) that is subject to scaling, clipping, and quantization to produce a vector of 8-bit output activations~\cite{burrello2020dory}. In the IMA, the intermediate representation is an analog current, while scaling, clipping and quantization are performed directly by the bitline ADCs by setting appropriate current limits. Recent work has shown that it is possible to achieve software-equivalent classification accuracy using this approach \cite{Y2020joshiNatComm}.

\subsection{Heterogeneous Cluster}

Highly parallelizable workloads such as DNNs are a perfect fit for high core count heterogeneous systems that can integrate specialized accelerators. We based our work on the PULP cluster~\cite{pullini2019mr} which structure can be seen at Fig.~\ref{fig:ima_subsystem} on the left. The cluster incorporates 8 RISC-V cores who share a single-cycle latency, word interleaved data memory called tightly coupled data memory (TCDM), or referred to as L1 memory. The cores are enhanced with a custom ISA extension called Xpulp~\cite{gautschi2017near} that aims to accelerate arithmetic intensive kernels. The work can be offloaded to accelerators (IMA in this case) by accessing the internal control register file via peripheral interconnect and programming them based on the workload.

\section{IMA Subsystem Architecture}
\label{sec:execution_model}

\begin{figure}[t]
\centerline{\includegraphics[width=0.5\textwidth]{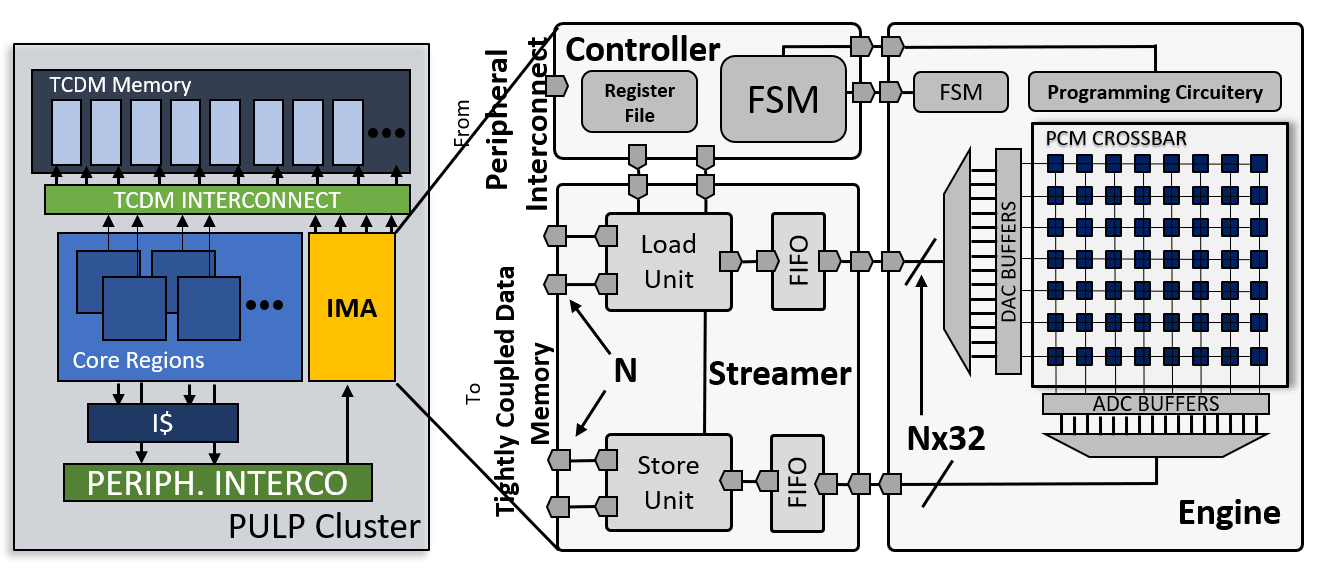}}
\caption{IMA Subsystem in a PULP cluster. The subsystem can be configured at design time to support a total of $2\times N$ 32-bit ports towards TCDM, evenly split between load and store units connected directly to the PCM array by means of DACs and ADCs, respectively.}
\label{fig:ima_subsystem}
\vspace{-5mm}
\end{figure}

The IMA exposes a control and a data interface towards the rest of the cluster based on a standardized Hardware Processing Engine (HWPE) interface~\footnote{https://hwpe-doc.readthedocs.io/en/latest/}. 
%Through the control interface, which is attached to the peripheral interconnect of the cluster, cores can directly access a set of control registers. 
The data interface employs a direct connection with TCDM memory, composed of 16 parallel 32-bit banks in this work, through the same interconnect used by cores. The number of master ports is a design-time parameter $N_{port}$ that can be chosen depending on the required bandwidth -- as we show in Section~\ref{sec:ima_std_conv},~\ref{sec:bottlenecs_res}.

In Fig.~\ref{fig:ima_subsystem}, we show a detailed view of the IMA subsystem. The accelerator is composed of three main blocks. The \textit{controller} includes the register file and the internal FSM coordinating the other blocks. The \textit{engine} contains both the digital and analog parts of the IMA datapath. The digital part is composed of buffers for ADCs and DACs and control circuitry; the analog core encloses all the PCM devices (including PCM programming circuitry), as well as the ADCs and DACs themselves. The \textit{streamer} block contains the address generators for memory transactions, implements the request protocols towards the TCDM, realigns data, and takes care of contentions. The address generators are capable of three-dimensional stridden access. Data coming from $N_{port}$ 32-bit TCDM ports are merged into a unique stream of data using a simple ready/valid handshake, which is fed to the engine. Conversely, data streams coming from the engine towards the TCDM memory are split in $N_{port}$ 32-bit TCDM accesses.

% \subsection{IMA configuration and execution model}
The configuration sequence of the IMA starts when a core acquires a lock over the accelerator by reading a special \textsc{acquire} register through the peripheral control interface.
% The accelerator can be configured via the peripheral interconnect, before any interactions the IMA has to be acquired by writing to the acquire address.
% Once acquired the accelerator will be in a busy state and only the core that got the access can write to the internal registers.
After that, the core can interact with the IMA by: programming the PCM devices with the weights of one or multiple layers; reading the conductance value of a PCM device; configuring a \textit{job} by setting the address of input and output data in TCDM and the ADC configuration; when the configuration is over the job can be started by writing to a special \textsc{trigger} register.
To minimize IMA configuration and synchronization overhead, multiple jobs can be pipelined by setting the register file with the correct strides. Thus, a whole layer can be executed with only one configuration phase.
%\textbf{Include tiny pseudo-code and profile of the computation phases.}
% 
% 
%\textbf{Missing detail on mapping strategy for DNNs.}
% The phases just described are managed by an engine FSM internal to the engine which is denominated as the following: Stream-in, Computation, Stream-out.

\begin{figure}[t]
\centerline{\includegraphics[width=0.45\textwidth]{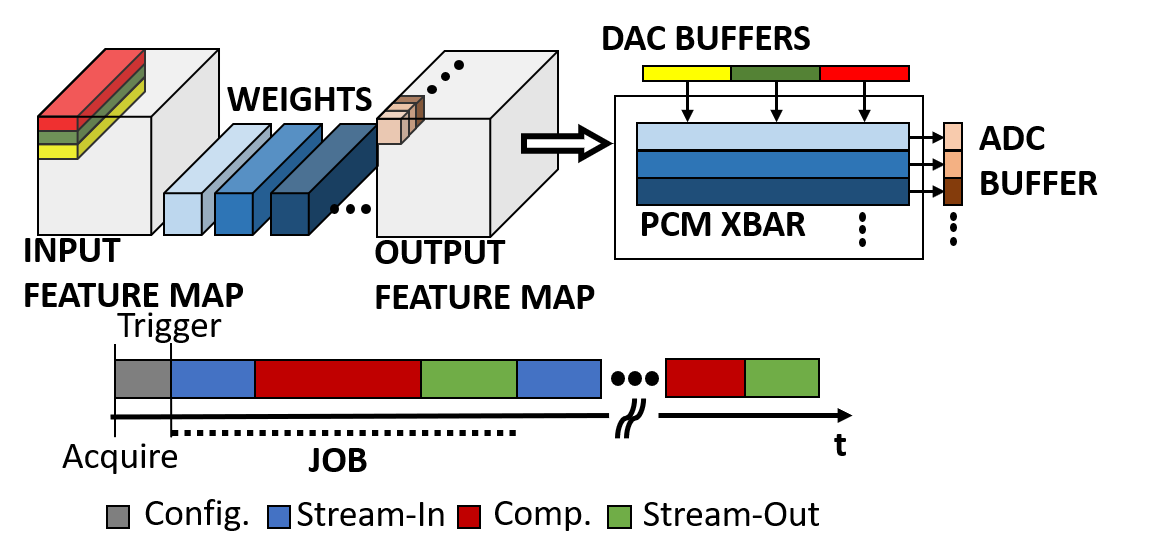}}
\caption{IMA mapping of standard convolutions on the PCM crossbar. Below a timeline of the execution model.}
\label{fig:ima_map_plus_timeline}
\vspace{-5mm}
\end{figure}

% \subsubsection{IMA execution model}
The IMA works on input data stored in L1 with the HWC format, i.e., with consecutive data elements encoding pixels that are adjacents in the channel dimension.
% We remap all layers that are executed on the IMA to matrix multiplications that can be executed on the PCM array; we focus on 
The execution of a job is divided into three phases: \textsc{streamin}: fetch data from the TCDM that is then streamed to the engine's internal DACs buffers; \textsc{computation}: analog computation on the crossbar and writing of the ADCs buffers; \textsc{streamout}: stream data from buffers back to the TCDM. In Fig.~\ref{fig:ima_map_plus_timeline}, we show how a CNN layer is mapped into the IMA and how the computational timeline is executed.
For a standard convolutional layer, the \textsc{streamin} phase also includes a virtual IM2COL transformation~\cite{garofalo2020pulp}, which is performed directly by the streamers, enabling to remap all computation supported by the IMA to matrix-vector products of the form discussed in Section~\ref{sec:in-mem_comp}.
As a consequence, the PCM array computes $C_{out}$ output feature maps from a complete input volume of $C_{in}\times K\times K$ pixels in a single operation, where $C_{in,out}$ indicate the number of channels and $K$ is the filter size.

\section{Results \& MobileNetV2 Case Study}

\subsubsection{Experimental Setup}
Our results are obtained by synthesizing the cluster while varying the number of TCDM master ports in the accelerator, using Synopsys Design Compiler to target the GlobalFoundries 22nm FDX technology (SSG corner @ 0.59V and 250 MHz).
% We characterized two scenarios for performance and power: software-based execution; the second one is the power consumed while the cluster was executing a CNN layer on the IMA.
% The tool used for Synthesis is Synopsys Design Compiler, using GF22FDX libraries. The operating conditions are SSG  constraints on the clock.
For power analysis, we used Synopsys PrimeTime with typical corner with 0.65V at 25°C, with switching activity back-annotation from post-synthesis simulation. The IMA model for performance and power are estimated from~\cite{hermesCor}; each PCM device has an area of 18.2 $\mu m^2$ and a single array operation takes 70ns. We assume that the PCM array is properly sized to fit all weights. All results are reported relative to the cluster frequency of 250MHz and using the convention 1 MAC = 2 OPs.

\subsubsection{Baseline IMA performance}
\label{sec:ima_std_conv}

\begin{figure}[t]
\centerline{\includegraphics[width=0.6\textwidth]{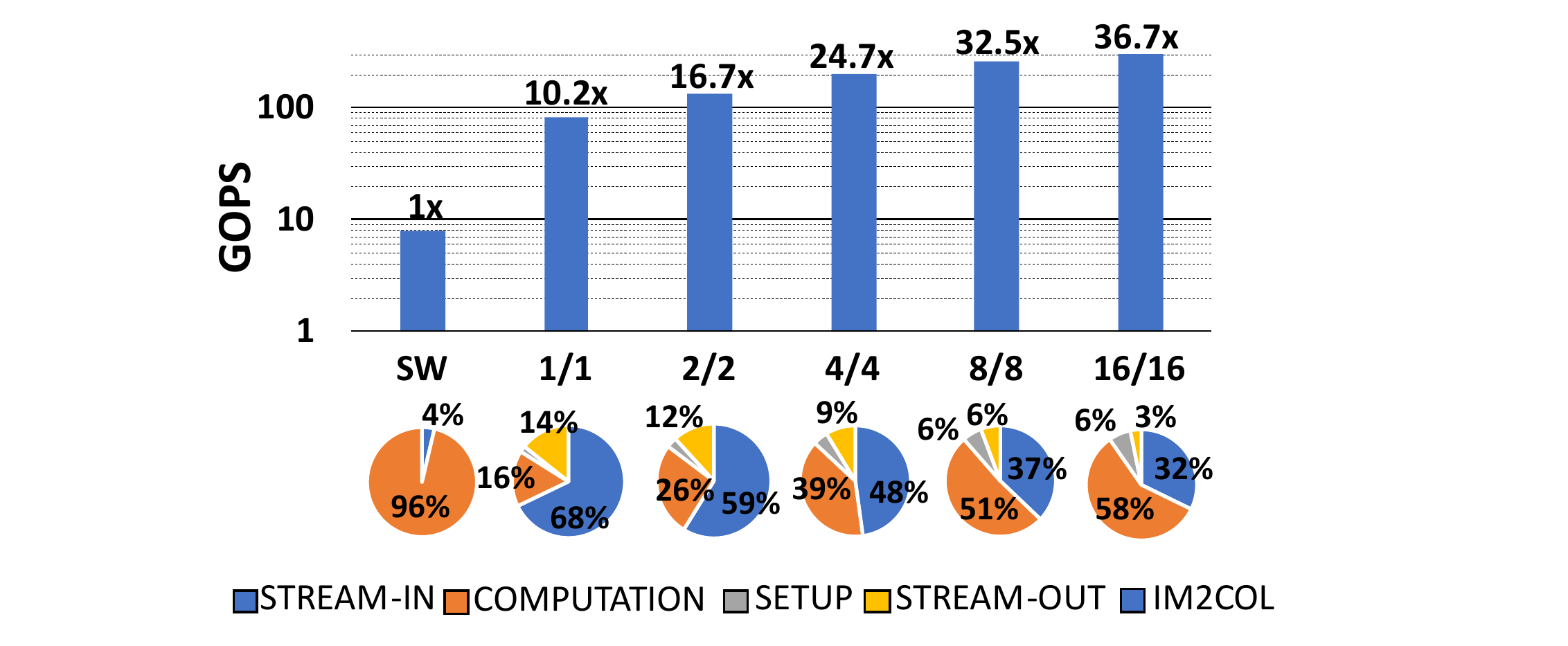}}
\caption{Performance on standard convolution: Software execution (SW) vs IMA acceleration (frequency = 250MHz). The N/N configuration indicates the number of load and store master ports (32-bit each), respectively.}
\label{fig:ima_opt}
\vspace{-5mm}
\end{figure}

% \subsection{IMA results on standard convolutions}
% %  change to something like IMA perf vs TCDM bw ?
% \label{sec:ima_std_conv}
% %\textbf{Maybe move to results?}
The peak theoretical performance of the PCM array is limited only by its size: for a 100x100 array, for example, 286 GOPS are achieved; for a 1000x1000 array, this would be 28.6 TOPS, etc. However, real performance is limited by \textit{i)} array utilization; \textit{ii)} how fast activation data is streamed in/out of the IMA. To assess the IMA's performance in a realistic baseline case, we used a standard convolutional layer with 3x3 filter, with 16x16 output size, 32 input channels, and 64 output channels ($\sim$4.7 MMAC).
% 
% which can be seen in Fig.~\ref{fig:ima_opt}.
% IMA presents multiple instances where \textsc{{x/y}} indicates the number of TCDM ports configurations for input and output respectively.
% The ports have been kept symmetrical since the amount of bandwidth in input or output depends on the layer executed (in this example emphasis was mostly on stream-in).
Fig.~\ref{fig:ima_opt} shows the IMA performance obtained while sweeping the number of load/store ports from 1/1 to 16/16, compared with a pure software execution on the 8 cores using PULP-NN~\cite{garofalo2020pulp} which achieves more than 55\% utilization of the SIMD MAC units in the cores.
% 
%We notice that on standard convolutions, the main limit to the IMA performance is given by the speed at which the data can be streamed in. 
In this case, a significant portion of the execution time is due to data stream-in which can be reduced by increasing the bandwidth to the IMA.
The speed-up ranges from 10.2$\times$ on \textsc{1/1} configuration up to 36.7$\times$ when \textsc{16/16} is used.
% On the core side, most of the time is spent on executing the inner loop of the kernel while 20\% of the cycles is consumed on data marshaling (IM2COL phase, which is not required on IMA thanks to the 3D transfers of the streamers). The remaining contribution on fig~\ref{fig:ima_opt} have the following meaning: \textit{stream-in overhead} includes contentions, fifos delays, data alignment; \textit{SETUP}, which is the number of cycles required to update addresses, enabling/disabling the PCM devices between jobs.

\subsubsection{Case Study: MobilenetV2}

\begin{figure}[t!]
\centerline{\includegraphics[width=0.45\textwidth]{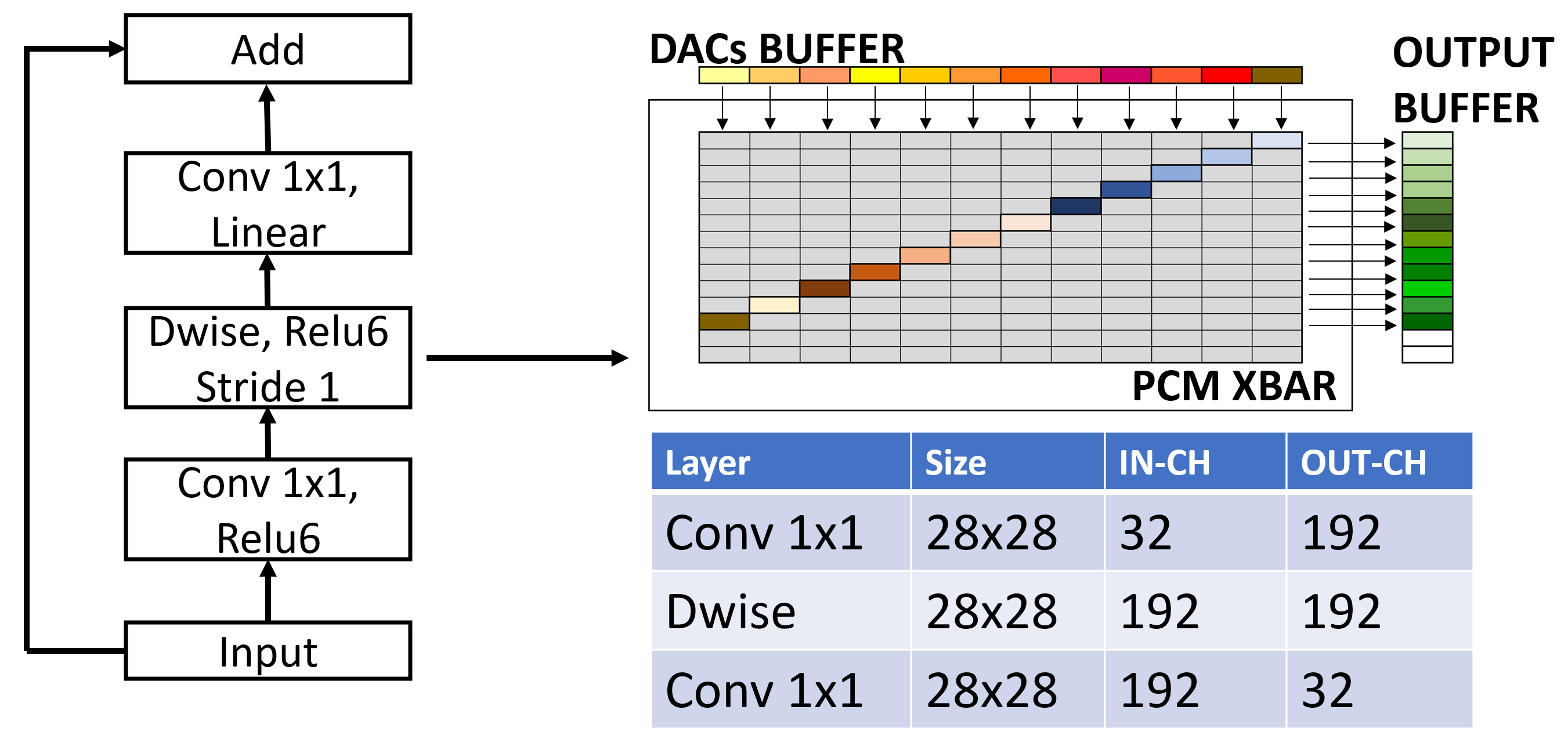}}
\caption{Components of MobilenetV2 bottleneck block with stride = 1 and mapping structure in the PCM crossbar for depthwise layers. All the gray rectangles are padding required for computing more than 1 channel per job.}
\label{fig:bottleneck}
\vspace{-5mm}
\end{figure}

To highlight the advantages and trade-offs of IMC on a realistic use case for extreme edge computing, we selected MobileNetV2, a widely used DNN benchmark constructed as a deep stack of units called \textit{BottleNecks}.
%\textit{
%\textbf{This text can be cut if space is scarce.}
%All BottleNecks share the same structure but differ from each other in terms of %number of in/out channels and spatial stride; their operation is based on a 1x1 %convolution used to expand tensors channel-wise by a factor of 6; a 3x3 depthwise %layer performing separate transformations on each channel; and another 1x1 %convolution compressing back the number of channels.
%Finally, the input of the BottleNeck is added to the output by means of a bypass %connection.
%}
For this analysis, we focus on the BottleNeck configuration that is shown in Fig.~\ref{fig:bottleneck}; the configuration is chosen so to fit the on-cluster TCDM (512 kB) without requiring any activation data tiling~\cite{burrello2020dory}, which is beyond the scope of this work.\footnote{We note that data tiling will further decrease energy efficiency.}
% The Bottleneck is made of several sub-layers where the first is a \textit{conv 1x1} that expands the number of channels by 6, followed by a depthwise layer and a pointwise that compress back the number of channels. The first and second layer have a \textit{RELU6} as activation function while the pointwise is only followed by a residual block that sums the output of the pointwise with the input of the bottleneck.
% In this work we consider a TCDM large enough (512 KB) to accommodate input and output features maps. Bottlenecks deeper in the network are significantly larger than the one tested and would not fit all in the TCDM requiring tiling. Studying bottlenecks that fit all in TCDM shows the maximum performance achievable with the system and the minimization of the tiling overhead is out of the scope of this work.

All layers in the BottleNeck can be mapped on the IMA. For the 1x1 convolutional layers, the mapping is direct as explained in Section~\ref{sec:execution_model}, exploiting their high-level of channel parallelism. However, in the 3x3 depthwise layer each output channel depends only on a single input channel. This fact means that optimizing at the same time the array utilization and the execution performance is not possible.

%The three layers of the bottleneck can be mapped to the IMA PCM devices directly but can be efficiently done only in convolution where the output also depends on the value across the input channels, which does not happen in the depthwise layer of the bottleneck: each channel output depends only on the values of the input at the same channel depth. This means that the input from other channels cannot be used to compute the output of the current channel requiring padding (as shown on the lower right side of  fig\ref{fig:bottleneck}). 

A $K\times K$ depthwise layer with $C$ in/out channels can be mapped as a standard layer with all weights out of a diagonal set to 0, as shown in Fig.~\ref{fig:bottleneck}. This means that out of $K^2 \times C^2$ crossbar locations, only $K^2\times C$ are useful, leading to low utilization of the array. On the other hand, the depthwise can be split in separate jobs for better array utilization, but this leads to a smaller amount of operations per job, reducing performance. The total number of crossbar elements required is in general given by $N_{xbar} = K^2 \times C \times C_{job}$, where $C_{job}$ is the number of channels per job. For a MobileNet-V2, full throughput for all Bottlenecks would require a 23$\times$ larger array than what simply counting the number of parameters would suggest. This result stands even if the number of depthwise parameters is just $\sim$4\% of the total number of weights.
% Clearly, this approach is not feasible and fixing the number of output channels computed per job will allow packing weights into devices more densely, and thus provide a trade-off between PCM devices and throughput.
In this work, we considered $C_{job}=8$ and $16$ as reasonable trade-off configurations, which translates to an increase of 25\% and 54\% in the number of devices respectively. These are indicated as \textsc{ima8} and \textsc{ima16}, respectively, in the following sections.
%The throughput of depthwise layers is also hampered by the output structure from the previous layer, which is stored as HWC, meaning the feature map is stored channel first, then width and last height. Depthwise layers executed on IMA require CHW (width first) structure which translates into higher overhead for fetching the input feature map due to data not being contiguous in memory. In this case, 9 requests to memory are necessary given the 3 by 3 kernel filter even though there is enough bandwidth to fetch everything in one request. %Depending on the number of fixed output chosen, 16 or 8 different channels are fetched from memory on each memory transaction and are then offset internally.
% 

\begin{figure*}[t]
\centerline{\includegraphics[width=\textwidth]{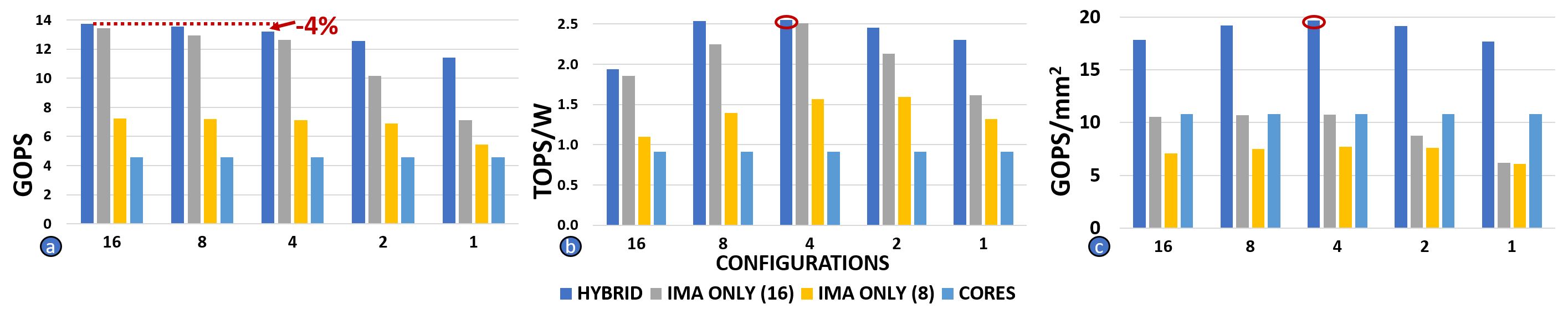}}
\caption{Performance of several design implementations of the MobileNetV2 bottleneck. a) Performance results measured in GOPS; b) Energy efficiency in TOPS/W; c) Area efficiency in GOPS/$mm^2$. Red circles on a) and b) indicate best results.}
\label{fig:res_compressed}
\vspace{-5mm}
\end{figure*}

An alternative solution supported by the heterogeneous cluster we propose is the parallel execution of the depthwise layer via software~\cite{garofalo2020pulp} on the 8 RISC-V cores of the cluster, intermixed with IMA-based execution of 1x1 layers. This configuration, which is reported as \textsc{hybrid}, requires the parameters from the depthwise layer to be stored in memory instead of IMA which we consider a reasonable trade-off since those parameters account only for ~4\% of the total weights.
% \subsubsection{MobileNetV2 Bottlenecks Results}

\begin{figure}[t!]
\centerline{\includegraphics[width=0.45\textwidth]{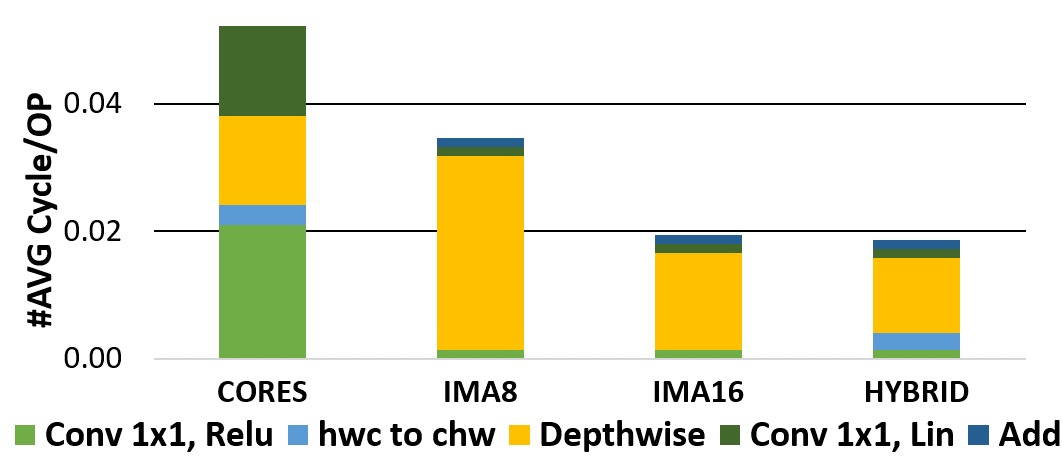}}
\caption{Impact on performance of the various Bottleneck phases (hwc to chw only needed on SW variants for depthwise). Results are taken using 4/4 port configurations at 250 MHz.}
\label{fig:totCycles}
\vspace{-5mm}
\end{figure}

The performance results in this section are from the Bottleneck with sizes reported in Fig.~\ref{fig:bottleneck} sweeping across 1 to 16 ports for stream-in and out. In Figs.~\ref{fig:res_compressed} (a-c) we can see how the benefits of adding TCDM master ports start to fall off after 4/4: the depthwise layer dominates the number of cycles (see Fig.~\ref{fig:totCycles}) and increasing ports doesn't render as sizeable an effect as shown in Fig.~\ref{fig:ima_opt}.
In particular, for the \textsc{hybrid} solution, increasing bandwidth toward IMA with more ports does not influence the depthwise execution.
In \textsc{ima16} configuration the bandwidth for depthwise saturates when all the channels can be fetched in one cycle: 4 TCDM ports of 4 bytes each are enough; going over only benefits 1x1 convolutions.
The same reasoning can be applied to 8 channels per job, where 2 ports are sufficient.

Thus, the importance of the depthwise layer in the Bottleneck drives the total improvement when using the IMA down to $\sim$3$\times$ the software implementation (down from $\sim$36$\times$ on standard convolutions).
Overall, the \textsc{hybrid} configuration stands out as the fastest: this is because even in the \textsc{ima16} configuration, the depthwise layer is slower than in software, as can be seen in Fig.~\ref{fig:totCycles}.
Similar considerations can be made with respect to energy efficiency, noticing that adding more ports than necessary reduces energy efficiency with respect to the peak at 4 (\textsc{hybrid}/\textsc{ima16}) or 2 ports (\textsc{ima8}), as it puts more pressure on the memory system.

To put in perspective the cost of increasing the throughput using IMA, the area efficiency reported in Fig~\ref{fig:res_compressed}(c) is relative to the effective area of the PCM arrays utilized to implement the Bottleneck (including padding). The \textsc{hybrid} solution has the best result requiring $\sim$3.25$\times$ and 2.13$\times$ smaller PCM area for the same bottleneck when compared to \textsc{ima16} and \textsc{ima8}, respectively. Considering also the area of the cluster itself, we obtain 1.82$\times$ and 2.56$\times$ better GOPS/$mm^2$, respectively.

% The fig~\ref{fig:res_compressed}(b) shows energy efficiency in TOPS/W. Here we see how the peak of efficiency is with 4 TCDM ports for the hybrid and IMA-16, while is 2 when looking at IMA-8. Scaling the number of TCDM master ports increases both the interconnect and IMA subsystem size in both area and power consumptions. Also, the higher bandwidth puts more pressure on the memory system rising his power consumption. The latter has a bigger impact in terms of power consumption than performance benefits hence reducing the energy efficiency.

\label{sec:bottlenecs_res}

If we compare the described \textsc{hybrid} implementation for MobileNetV2 bottleneck with a s.o.a. heterogeneous system where the IMA is integrated through a loosely-coupled 32-bit AXI bus controlled by a tiny processor \cite{jia2020programmable}, our proposed solution performs $\sim$45$\times$ better. First, our core is $\sim$10$\times$ faster on depthwise convolutions on a per-core basis. If we expand the picture to the entire cluster we would look at another $\sim$7$\times$ improvement factor. Finally, the loosely coupled integration of the IMA through a 32-bit system-bus \cite{jia2020programmable} forms a huge performance bottleneck even for standard convolutions (see Fig.~\ref{fig:ima_opt}). Our high-bandwidth tightly-coupled interconnect scheme allows delivering to the IMA the required bandwidth in a scalable way with low communication latency with the cores, improving by $\sim$3.5$\times$ the IMA performance over \cite{jia2020programmable}, making the proposed hybrid solution viable, and paving the way for a new generation of architectures exploiting synergies between analog and digital computing. Finally, the results of our exploration in Fig.\ref{fig:totCycles} suggests that further gains in the order of 10$\times$ can be achieved by extending these clusters with specialized digital accelerators better tuned for key functions of modern DNNs.

%This originates from the fact that, increasing the number of ports has a bigger impact on power consumption than on performance.
% In the fig~\ref{fig:res_compressed}(c) we report the GOPS/$mm^2$, a metric for area efficiency.

% Last graph in fig~\ref{fig:totCycles} we show the number of cycles it takes to complete the bottleneck with 4 TCDM port configurations at 250 MHz. The components are split into sections as shown in \ref{fig:bottleneck}. Dominating the computation time when using the IMA is depthwise, increasing the number of output channels reduces the number of jobs required, and it scales quite linearly given the fact that the output of the PCM crossbar requires a fixed time to evaluate the output. When looking at the convolutions 1x1, the amount of speed-up achieved is comparable with the fig~\ref{fig:ima_opt} when using the software version as reference.

\section{Conclusion \& Discussion}
In this work, we integrated an In-Memory Accelerator (IMA) into a cluster of 8 RISC-V cores.
As expected, the IMA boosts performance in standard convolutions by a significant factor (up to 36$\times$ when compared to an 8-cores cluster in our experiments).
However, our results also show that the inflexible Matrix-Vector product paradigm imposed by IMAs requires some mitigation on the architectural side. % if one aims at extracting optimal energy efficiency.
This observation strongly motivates our choice to couple a highly efficient IMA with a highly flexible cluster of cores.
In fact, even a relatively simple Bottleneck layer from a MobileNetV2 includes blocks that are not well-mapped to the IMA, specifically, depthwise separable convolutions.
We show several possible mappings trading off area and performance, demonstrating that executing depthwise layers directly in the cores yields up to 2.56$\times$ better area efficiency without overhead in performance and energy.
The heterogeneous system achieves 13.2 GOPS, 19.7 GOPS/$mm^2$ and 2.55 TOPS/W on a \textsc{4/4} configuration that is competitive with declared metrics from s.o.a. academic~\cite{jia2020programmable} and commercial systems~\cite{mythicHC2018}.
We argue that enhanced architectural heterogeneity is the key to fully exploit the potential of IMC architectures by offsetting their current limitations.
Our future work includes further extending heterogeneous clusters with digital accelerators tuned to key kernels that are not well suited to IMC, such as depthwise layers, nearing the 100 TOPS/W targets in real-world DNN inference.

\bibliographystyle{IEEEtran}
\bibliography{ref}

\end{document}